\documentclass[11pt]{article}
\pdfoutput=1

\usepackage{graphics, color,soul}
\usepackage{graphicx}
\usepackage{amssymb}

\usepackage{lmodern,mathtools}

\usepackage{booktabs}
\usepackage[english]{babel}
\usepackage{amsmath,amssymb,amsbsy,amstext, amsthm, simplewick}
\usepackage{hyperref}
\usepackage{tikz}
\usepackage{cite}

\usetikzlibrary{decorations.pathmorphing,shapes.misc}
\tikzset{snake it/.style={decorate, decoration=snake}}
\tikzset{cross/.style={cross out, draw=black, minimum size=2*(#1-\pgflinewidth), inner sep=0pt, outer sep=0pt},
cross/.default={1pt}}

\usepackage{amsfonts}
\usepackage{amssymb}
\usepackage{upgreek}
\usepackage{simplewick}
 \usepackage{exscale,relsize}
\usepackage{mathtools}
\usepackage{comment}

\usepackage[margin=1cm,labelfont={sf,bf,scriptsize},textfont={sf,scriptsize}]{caption}

\usepackage{colortbl}
\definecolor{lightgreen}{cmyk}{0.2, 0, 0.2, 0.2}
\definecolor{lightgray}{cmyk}{0.1,0.2,0,0.1}
\definecolor{lightgray2}{cmyk}{0.1,0.1,0,0.1}

\makeatletter
\newlength{\apb@width}
\newcommand{\autoparbox}[2][c]{\settowidth{\apb@width}{#2}\parbox[#1]{\apb@width}{#2}}

\makeatother

\numberwithin{equation}{section}

\def\beq{\begin{equation}}
\def\eeq{\end{equation}}

\def\bea{\begin{eqnarray}}
\def\eea{\end{eqnarray}}

\def\beq{\begin{equation}}
\def\eeq{\end{equation}}
\def\be{\begin{equation}}
\def\ee{\end{equation}}
\def\bea{\begin{eqnarray}}
\def\eea{\end{eqnarray}}

\def\0{{\vec{0}}}

\DeclareRobustCommand{\SkipTocEntry}[4]{}

\def\beq{\begin{equation}}
\def\eeq{\end{equation}}

\def\ba#1\ea{\begin{align}#1\end{align}}
\def\bg#1\eg{\begin{gather}#1\end{gather}}
\newcommand{\bseq}{\begin{subequations}}
\newcommand{\eseq}{\end{subequations}}

\DeclareSymbolFont{extraup}{U}{zavm}{m}{n}
\DeclareMathSymbol{\varheart}{\mathalpha}{extraup}{86}
\DeclareMathSymbol{\vardiamond}{\mathalpha}{extraup}{87}

\def\({\left(}
\def\){\right)}
\def\[{\left[}
\def\]{\right]}

\begin{document}

\begin{titlepage}

\setcounter{page}{1} \baselineskip=15.5pt \thispagestyle{empty}

\vbox{\baselineskip14pt
%\hbox{hep-th/0000000}
}
{~~~~~~~~~~~~~~~~~~~~~~~~~~~~~~~~~~~~
~~~~~~~~~~~~~~~~~~~~~~~~~~~~~~~~~~
~~~~~~~~~~~ }

\bigskip\
%\hbox{CALT-TH-2019--031}
\vspace{1cm}
\begin{center}
{\fontsize{19}{36}\selectfont  
{
A new branch of inflationary speed limits
%Log Inflation
}
%SITP-18/02
}
\end{center}

\vspace{0.6cm}

\begin{center}
Dayshon Mathis, Alexandros Mousatov, George Panagopoulos, Eva Silverstein
\end{center}

%\vspace{0.2cm}

\begin{center}
\vskip 8pt

\textsl{
\emph{Stanford Institute for Theoretical Physics, Stanford University, Stanford, CA 94306, USA}}
%\vskip 7pt
%\textsl{\emph{$^2$Walter Burke Institute for Theoretical Physics, California Institute of Technology, Pasadena, CA 91125, USA}}
%\vskip 7pt
%\textsl{\emph{$^3$Institute for Quantum Information and Matter, California Institute of Technology, Pasadena, CA 91125, USA}}
%\vskip 7pt
%\textsl{ \emph{$^4$Centro At\'omico Bariloche and CONICET, Bariloche, Argentina}}

\end{center}

\vspace{0.5cm}
\hrule \vspace{0.1cm}
\vspace{0.2cm}
{ \noindent \textbf{Abstract}
\vspace{0.3cm}

We present a new mechanism for inflation which exhibits a speed limit on scalar motion, generating accelerated expansion even on a steep potential.  This arises from explicitly integrating out the short modes of additional fields coupled to the inflaton $\phi$ via a dimension six operator, yielding an expression for the effective action which includes a nontrivial (logarithmic) function of $(\partial\phi)^2$. The speed limit appears at the branch cut of this logarithm arising in a large flavor expansion, similarly to the square root branch cut in DBI inflation arising in a large color expansion. Finally, we describe observational constraints on the parameters of this model.

\vspace{0.4cm}

 \hrule

\vspace{0.6cm}}
\end{titlepage}

\tableofcontents

%%%%%%%%%%%%%%%%%%%%%%%%%%%%%%%%%%%%%%%%%%%%%%%
%%%%%%%%%%%%%%%%%%%%%%%%%%%%%%%%%%%%%%%%%%%%%%%
%%%%%%%%%%%%%%%%%%%%%%%%%%%%%%%%%%%%%%%%%%%%%%%
\section{Introduction}\label{sec:intro}

Early universe inflation is a mature field, combining theory and observation to account for the origin of structure in the universe, in the process constraining detailed physics pertaining nearly 14 billion years ago.  It is of great interest to pursue this to the fullest extent possible.

One of the intermediate developments in the field was the realization that single-field inflation may proceed even on a steep potential, given appropriate self-interactions of the inflaton field of the sort introduced in \cite{Armend_riz_Pic_n_1999}.  
The earliest example of this was DBI inflation \cite{Alishahiha:2004eh, Silverstein:2003hf}, featuring a speed limit on scalar field motion tied to a square root branch cut in the effective action.  In the context of the AdS/CFT correspondence, such an action arises from an infinite sequence of powers of first derivatives of the field, arising from integrating out off-diagonal Yang-Mills fields in the large-$N_{color}$ theory (as inferred from the gravity side of the duality).   

In this work, we present a new example of this phenomenon, with a logarithmic branch cut that arises from a summation accessible via a large $N_{flavor}$ expansion.  It proceeds from a formula derived in \cite{Panagopoulos:2019ail}\ (and anticipated in a different context in \cite{Tolley:2009fg}), with a distinct choice of signs.  We work at the level of effective quantum field theory, with explicit calculation of the inflaton effective action.   

The early examples contributed to a more systematic understanding of inflationary theory and associated non-Gaussian signatures, which includes model-independent treatments \cite{Senatore_2016}.  This general class of inflationary scenarios features a small sound speed of perturbations, with current observational constraints from the cosmic microwave background \cite{Akrami:2019izv}\ partially constraining the parameter space (\ref{csbound}).  Large-scale structure measurements could cover some fraction of the remaining region.  
The effect could be rather general even among single-field inflationary models, but since it involves a resummation of kinetic effects there are not many specific examples in the literature.   However, interesting realizations appear in \cite{Tolley_2010}\cite{Baumann:2011su} and others such as the recent work \cite{Garcia-Saenz:2018ifx}, and related mechanisms such as \cite{Green:2009ds}\ also produce the equilateral signal.  
This motivates additional examples, such as the new class of models we present in this work.  

This paper is organized as follows.  We first derive the effective action for the inflaton field.  Then we analyze the dynamics at the homogenous level in sections 3 and 4.  First, in \S3 we work in Minkowski spacetime and use energy conservation to immediately deduce the speed limit on scalar motion before describing more details of the evolution.  We then incorporate gravity in \S4, starting with a simple derivation of the inflationary dynamics.  Finally, in \S5 we summarize observational constraints on the parameters of our model.  

\section{Deriving the inflaton effective action}

We start from the classical Lagrangian 
\begin{equation}\label{classS}
\mathcal{S}=\int \sqrt{-g}\left\{\frac{1}{2}(\partial\phi)^2\left(1+\delta K_\phi+\frac{\vec\chi^2}{M_*^2}\right)+\frac{1}{2}(\partial\vec\chi)^2-V(\phi) - V(\vec\chi) \right\} \end{equation}
depending on the inflaton $\phi$ and an $N_f$-component vector of additional scalar fields $\vec \chi$.  We have allowed for a constant parameter $\delta K_\phi$ in the kinetic term; this is useful to keep track of but will play a limited role in our analysis.    
We work in mostly plus signature, with $(\partial\phi)^2=-g^{\mu\nu}\partial_\mu\phi\partial_\nu \phi$.  

For simplicity, we work in a region of the potential landscape of this theory for which  $m_\chi^2 = |\partial_\chi^2 V| < H^2$.  As described in \cite{Panagopoulos:2019ail}, integrating out modes of $\chi$ shorter than $H^{-1}$  at the Gaussian level in an inflationary FRW background
\begin{equation}\label{metric}
g_{\mu\nu} dx^\mu dx^\nu=-dt^2+a(t)^2 d{\vec x}^2, ~~~~ a(t)\simeq e^{Ht}
\end{equation}
yields the effective action for $\phi$ 
\begin{eqnarray}\label{1PIintegral}
\Gamma_{1PI} = \int a^3 \Big\lbrace \frac{1}{2} \Big(1 + \delta K_\phi +\frac{\vec\chi^{~2}}{M_*^2}\Big)(\dot\phi^2-\frac{1}{a(t)^2}(\partial_{\vec x}\phi))^2 + \frac{1}{2}(\partial \vec\chi)^2 - V(\phi) - V(\vec\chi) \\
- \frac{N_f}{2}\int_{H}^{M_{UV}} \frac{d^4 k_E}{(2\pi)^4} \log \Big(1 - \frac{(\dot\phi^2-\frac{1}{a(t)^2}(\partial_{\vec x}\phi))^2}{M_*^2(k^2_E + m_\chi^2 - i\epsilon)}\Big) \Big\rbrace
~~~~{\text{plus IR threshold effects}} \nonumber
\end{eqnarray} 
\noindent in direct analogy to the Coleman-Weinberg calculation of the 1-loop effective potential \cite{Chen:2018cts, Coleman:1985rnk}.  This effect was considered in the interesting work \cite{Tolley:2009fg}, as a controllably small correction to a classical procedure deriving ``$P(X)"$ models (with $X=(\partial\phi)^2$) from a larger framework with nontrivial field-space geometry.  Both in \cite{Panagopoulos:2019ail}\ and in \cite{Tolley:2009fg}, the loop effect was taken as subdominant in the dynamics.  Here, we take the opposite approach and consider a large flavor regime where the one-loop contribution dominates.  

In our large-$N_f$ regime, 1-loop effects dominate.  Here, by threshold effects, we are referring to contributions arising from $\chi$ modes with physical momentum of order $H$, the infrared limit of the integral.  These will affect the details of the physics at this scale, but not the qualitative features of our model.  The reason is becoming clear at this point in our analysis from the second line of our expression (\ref{1PIintegral}):  at the homogeneous level, as we increase the squared field velocity $\dot\phi(t)^2$, it will hit the branch cut of the logarithm first at the scale $k_E \sim H$:  this scale will be robust against effects of the de Sitter geometry.  At this level, we may similarly neglect the $m_\chi^2$ contribution and we will drop it in the remainder of our analysis.  

The $\int d^4 k_E$ can be done explicitly.  Working at the level of spatially homogeneous fields, and neglecting $m_\chi^2$, we obtain for the last term of (\ref{1PIintegral}), including the $-$ sign, %{\color{red}GP: I think the penultimate term should have $- (M_{UV}^4 - \frac{\dot \phi^4}{M_*^4})$ in front}
\begin{equation}\label{1PIaction}
\frac{N_f}{64 \pi^2} \Big \lbrace (M^2_{UV}-H^2)\frac{\dot \phi^2}{M_*^2} + (H^4-\frac{\dot \phi^4}{M_*^4}) \log \Big(1-\frac{\dot \phi^2}{M_*^2 H^2}\Big) - (M_{UV}^4 - \frac{\dot \phi^4}{M_*^4}) \log \Big(1-\frac{\dot \phi^2}{M_*^2M_{UV}^2}\Big) +\frac{\dot\phi^4}{M_*^4}\log(\frac{M_{UV}^2}{H^2})\Big \rbrace
\end{equation}
The first term is absorbed into the coefficient of the $\dot \phi^2$ kinetic term, while for $\dot \phi \ll M_{UV}M_*$ 
%{\color{red} Shouldn't this be $\dot \phi \ll M_{UV}M_*$?} 
the penultimate term simplifies to
\begin{equation}
\frac{N_f}{64 \pi^2} \frac{M^2_{UV}}{M_*^2} \dot \phi^2
\end{equation}
so it similarly contributes to the $\dot \phi^2$ coefficient. 

At the two-derivative level, in order to analyze the dynamics, it is often most convenient to work with the canonically normalized scalar field. 
As we just computed, the effective Lagrangian contains the two derivative term 
%{\color{red} Brackets around $\dot\phi^2$? Also I think the first two terms should be divided by 2}
\begin{equation}\label{phdotsqterm}
\frac{1}{2}K_\phi\dot\phi^2 = \left(\frac{1}{2}(1+\delta K_\phi)+\frac{N_f}{64\pi^2}\frac{2 M_{UV}^2-H^2}{ M_*^2}\right) \dot\phi^2\equiv \frac{1}{2}\dot\phi_c^2
\end{equation}
where we defined the canonically normalized field $\phi_c$.  
However for now we will proceed with the action written in terms of $\phi$, ultimately seeking solutions where  logarithmic term dominates over the two-derivative kinetic term since it is responsible for the novel dynamics of our model.    

We will work in a simplifying regime which respects the parameter inequalities listed in \S\ref{sec:inequalities}.  In this regime, we can simplify the effective action for $\phi$ as follows:
\begin{equation}\label{Gammashort}
    \Gamma[\phi]\simeq \int a(t)^3\left\{\frac{1}{2}K_\phi\dot\phi^2+\frac{N_f}{64\pi^2}\Big(H^4-\frac{\dot \phi^4}{M_*^4}\Big) \log \Big(1-\frac{\dot \phi^2}{M_*^2 H^2}\Big) -V(\phi)\right\} 
\end{equation}
We will be interested in a regime where we are near the speed limit, $\dot\phi^2\lesssim H^2 M_*^2$, an inequality that will be enforced by the dynamics as we will see explicitly.  Although the logarithmic term in (\ref{Gammashort}) goes to zero in this limit, its contribution to the dynamics (the equations of motion, and also the Hamiltonian) becomes large, as we will derive in detail in the following sections.  In particular, the logarithmic branch cut will play a key role.

\section{Non-gravitational dynamics}

We will start by considering this theory in flat spacetime ($a(t)=1$), still denoting by $H$ the lower limit of integration over our $\vec\chi$ modes.  (This could arise effectively upon putting the theory in a box of size $H^{-1}$.)   In this situation, there is time translation symmetry and energy conservation, leading to a relatively simple analysis of the dynamics.

Starting from (\ref{Gammashort}), we obtain the field momentum as
\begin{align}\label{Pishort}
\Pi &= \frac{\partial \Gamma}{\partial \dot{\phi}}\nonumber\\
 &\simeq K_\phi\dot{\phi} - \frac{N_f}{32 \pi^2}\frac{\dot\phi}{M_*^2}(H^2+\frac{\dot \phi^2}{M_*^2})
 -\frac{N_f}{16\pi^2}\frac{\dot\phi^3}{M_*^4}\log(1-\frac{\dot \phi^2}{M_*^2 H^2})\nonumber\\
 &\to K_\phi HM_*-\frac{N_f}{16\pi^2}\frac{H^3}{M_*}\log(1-\frac{\dot \phi^2}{M_*^2 H^2}), ~~~~~~{\dot\phi\to HM_* }\nonumber\\
\end{align}

which gives near the speed limit 
\begin{equation}\label{Hamshort}
\mathcal{H}_{\text{eff}} =\Pi \dot\phi -\Gamma \simeq \frac{1}{2}K_\phi\dot{\phi}^2 - \frac{N_f}{16\pi^2}\frac{\dot\phi^4}{M_*^4}\log(1-\frac{\dot \phi^2}{M_*^2 H^2})+V(\phi), ~~~~\dot\phi\simeq HM_* \end{equation}
We note here that in this regime, $\dot\phi\simeq HM_*$, the logarithmic term in (\ref{Hamshort}) is positive.

In the non-gravitational setting, energy conservation applies and it will prevent this term from growing arbitrarily large, regardless of how steep the potential $V(\phi)$ is.  Before getting into any details, it is therefore manifest that the speed limit will be respected.

\begin{figure}
    \centering
    \includegraphics[scale=.1]{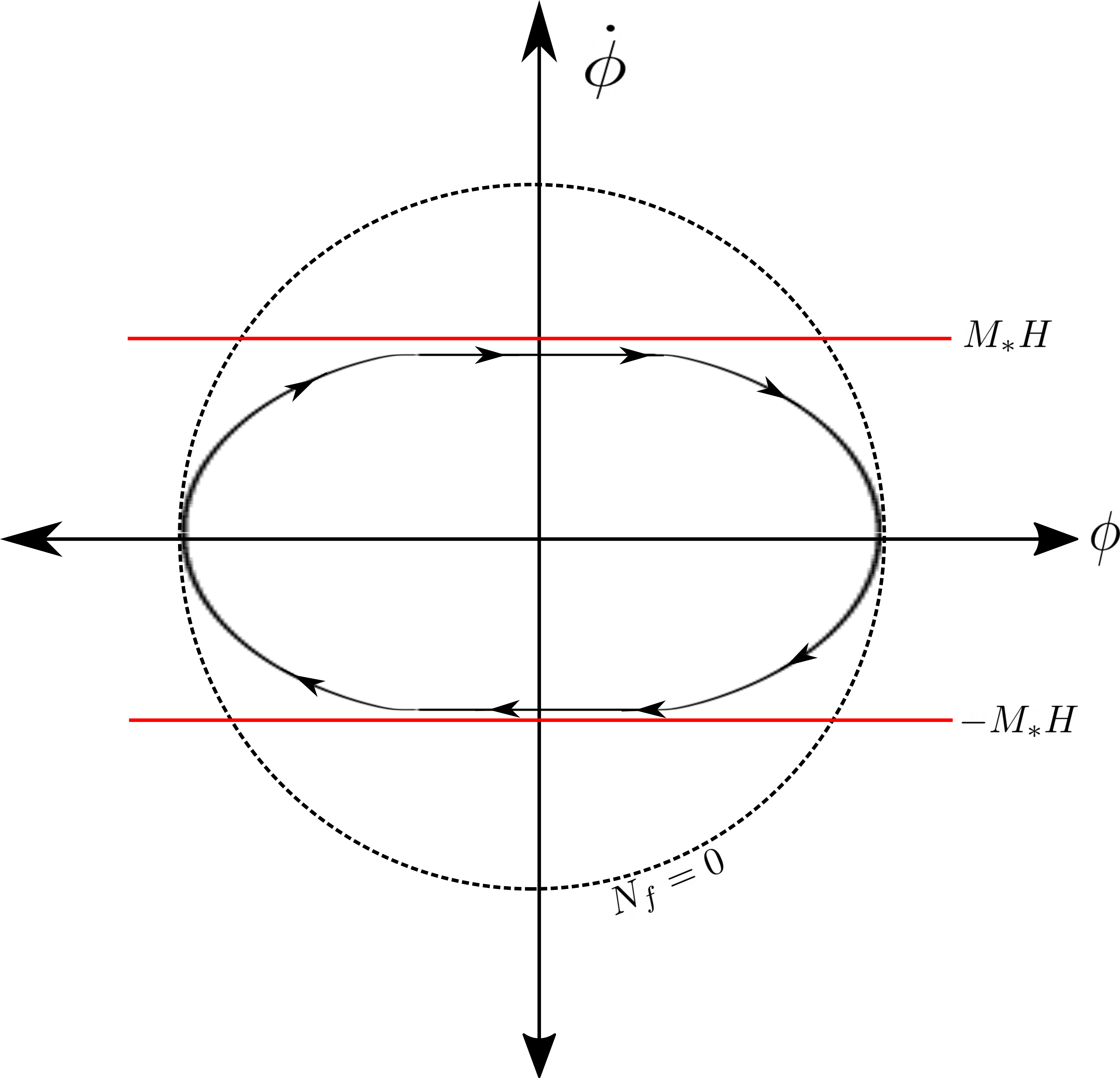}
    \caption{
    The phase diagram with finite $N_f$ will be limited by the speed limit shown by the red horizontal lines in the regime of parameters described in the text.}
\label{fig:phase}
\end{figure}

{
To analyze the problem further near the speed limit, we can write $\dot\phi(t) =H M_*(1-\delta(t))$ and $\phi(t) \approx \phi_0 + H M_* t$. Here, the negative quantity $\phi_0$ is the smallest value of $\phi$ where the approximation (\ref{Hamshort}) to the Hamiltonian is a good one. Then, the leading terms in the equation for energy conservation are
\begin{equation}
    -\frac{N_f H^4}{16\pi^2}\log(2\delta(t))+V(\phi_0 + H M_* t) = E
\end{equation}
where $E$ is an energy scale depending on the initial energy and $K_\phi$. This gives
\begin{equation}
    \delta(t) = \frac{1}{2}\exp\left(-\frac{16\pi^2}{N_f H^4} (E-V(\phi_0 + H M_* t))\right)
\end{equation}
Thus, in the non-gravitational case, $\dot\phi$ is exponentially close to the speed limit and stays there until the potential turns around. This is illustrated in Figure \ref{fig:phase}. We have also shown the case with no speed limit, or $N_f = 0$ for comparison. We will see in the next section that when gravity is reintroduced, $\dot\phi$ does not necessarily approach the speed limit exponentially; regardless,  the dynamics will support inflation on a potential too steep to inflate.  
}

%{\color{green}
%The conserved Hamiltonian density in terms of $\dot{\phi}$ is then
%\begin{equation}
%\mathcal{H}_{\text{eff}} = \Pi \dot\phi -\Gamma = \frac{1}{2}\dot{\phi}^2 + V(\phi) + \frac{N_f}{64\pi^2}\int^{k=M_{UV}}_{k=H}d\{k^4\log{(1-\frac{\dot{\phi}^2}{M_*^2 k^2})} + 3 \frac{\dot{\phi}^2}{M_*^2}(k^2 + \frac{\dot{\phi}^2}{M_*^2}\log{( k^2 -\frac{\dot{\phi}^2}{M_*^2})~)}\}
%\end{equation}
%The logarithm term is unbounded for finite values of $\dot{\phi}$, so if energy is conserved then momentum is bounded below $M_*H$, that is if $N_f \neq 0$. The relationship between $\phi$ and $\dot{\phi}$ with a constant Hamiltonian are shown on the phase diagram in figure \ref{fig:phase}.

%When $\phi$ is large and its rate is zero the slope of the potential will initiate motion. Other than the starting motion and confining the phase space the potential has no strict constraints and can satisfy slow roll conditions-- no steep potential required.

%The dominant terms in the equation of motion near the speed limit are
%\begin{equation}
%\frac{N_f H^3}{16\pi^2 M_*^3} \ddot{\phi} = -m^2 \phi (M_* H -\dot{\phi})
%\end{equation}
%The following time spent in that regime (assuming appropriate choice of initial conditions, $\phi_0$ and $C$) gives
%\begin{equation}
 %   t = \frac{1}{M_* H}\int^\phi_{\phi_0}\frac{d\xi}{1+W(Ce^{\lambda \xi^2})}
%\end{equation}
%with $\lambda = \frac{16\pi^2 m^2 M_*^2}{N_f H^4}$ and $W(x)$ is the Lambert function. This time is finite and so the trajectory for $\phi$ and $\dot{\phi}$ do not freeze near the speed limit in this case. For a total time $T$ spent near the speed limit the scalar field has magnitude
%\begin{equation}
 %   \phi = M_* H (T-t)(1+W(C e^{\lambda \phi_0^2}))
%\end{equation}
% and rate 
% \begin{equation}
%     \dot{\phi} = M_* H (1+W(C e^{\lambda \phi_0^2}))
% \end{equation}
% the Lambert function has argument such that its value is negative.
%}

\section{Inflationary Dynamics}

In this section, we reintroduce gravity and analyze the homogeneous cosmological evolution of our model.

\subsection{Parametric regime of inflation}\label{sec:inflationgeneral}  

Before getting into the details, we may establish inflationary dynamics rather simply as follows.  For this discussion, let us work with the case $K_\phi\simeq 1$, choosing $\delta K_\phi$ appropriately; we will later keep track of the more general parameter windows in the absence of such tuning, as collected in the appendix.    

Then, given
\begin{equation}\label{ineqM}
    M_*\ll M_P
\end{equation}
we see immediately that the speed limit ensures that the standard kinetic term is subdominant to the potential energy:
\begin{equation}\label{koneV}
    \dot\phi^2\simeq M_*^2 H^2\ll M_P^2 H^2 \sim V 
\end{equation}
where in the last step, we assumed that the potential altogether dominates in the Friedmann equation,
$H^2 \sim {\cal H}_{eff}/M_P^2$.  That depends on the subdominance of the logarithmic term in the energy density ${\cal H}_{eff}$ (\ref{Hamshort}), which evaluates near the speed limit to
\begin{equation}
    \frac{N_f}{16\pi^2}H^4\log(1-\frac{\dot\phi^2}{M_*^2 H^2})
\end{equation}
We would like to understand if this is also subdominant to $M_P^2 H^2$, so that the potential energy indeed dominates, driving inflation. In fact it is already clear that this is possible, since the dynamics determining how close $\dot\phi$ is to the speed limit is independent of $M_P$.  As long as we choose our field content such that  $\frac{N_f}{16\pi^2}$ also does not scale with $M_P$, it is guaranteed that the logarithmic term is also subdominant to the potential energy term, and our model will inflate.  

In the remainder of this section, we will spell this out in detail, indeed finding that the number of e-foldings is controlled by $M_P/M_*$.  In the final section we will summarize phenomenological predictions.      

\subsection{More detailed analysis of inflationary dynamics}\label{sec:inflationdetails}

We will be using the Hamilton-Jacobi method to find inflationary solutions to the Friedman equation
\begin{equation}
3 H^2 = \frac{1}{M_P^2} \mathcal{H}_{eff}
\end{equation}
where $\mathcal{H}_{eff} = \frac{\partial \mathcal{L}_{1PI} }{\partial \dot \phi} \dot \phi - \mathcal{L}_{1PI}$ is the energy density. By taking the time derivative of the above equation and using the equations of motion for $\phi$, we obtain the equation
\begin{equation}
6H\dot{H} = -\frac{1}{M_P^2}\left(3H  \frac{\partial \mathcal{L}_{1PI}}{\partial \dot \phi} \dot \phi + \frac{\partial \mathcal{L}_{1PI}}{\partial H} \dot H\right)
\end{equation}
Viewing $H$ as a function $\phi$ instead of time we arrive at
\begin{equation}\label{EoM}
    2M_P^2H' + \frac{N_f}{48\pi^2} H^2 H' \log(1-\frac{\dot\phi^2}{M_*^2 H^2}) = - \frac{\partial \mathcal{L}_{1PI}}{\partial \dot \phi}
\end{equation}
where the prime indicates a derivative with respect to $\phi$.

Our goal is to solve (\ref{EoM}) to find $\dot{\phi}(H, H')$, and then use this to express $\mathcal{H}_{eff} = \mathcal{H}_{eff}(H, H')$. This will reduce the Friedman equation to an ordinary differential equation that can be solved to obtain $H(\phi)$. At the end of the day, we will have to ensure that the solution is inflationary (i.e. $H^2 + \dot{H} > 0$) and consistent with any approximations we will make along the way.

To make the best use of our speed limit, we will assume that $\dot \phi \simeq M_* H$ to solve the equations, and at the end we will ensure that the assumption is consistent.
We can view our model at various levels: without entering into phenomenological constraints,  it will provide a mechanism for accelerated expansion; as we saw in \S\ref{sec:inflationgeneral}, this effect is rather general.
However, to analyze the details of the model, including a regime that satisfies observational constraints, it is useful to work in a subspace of the parameter space where the logarithmic term in (\ref{Gammashort}) dominates. Let us write $\dot{\phi} = H M_* (1-\delta)$ and expand at small $\delta$. 
In order to maintain consistency with observational constraints that are analyzed in the next section, we can assume that $\delta \ll 1$ but not necessarily that $\log(1/\delta) \gg 1$. In the regime where $\log(1/\delta) $ is of order 1, the dominance of the logarithmic term in (\ref{Gammashort}) requires tuning $K_\phi \simeq \frac{N_f H^2}{16\pi^2 M_*^2}$.  This is consistent with the inequalities we list in the appendix \S\ref{sec:inequalities}.
Given this, (\ref{EoM}) reduces to
\begin{equation}\label{SingularEoM}
\frac{N_f}{16 \pi^2} \frac{\dot \phi^3}{M_*^4} \log\Big( \frac{1}{1-\frac{\dot \phi^2}{M_*^2 H^2}}\Big) = -2 M_P^2 H'
\end{equation}
where we have also dropped the second term on the right hand side of (\ref{EoM}) as $M_P \gg H$ and $N_f \log(2\delta)$ is not many of orders magnitude larger than 1.

In our regime where $\delta\ll 1$, dropping all $O(\delta)$ terms we find that a $\log(2 \delta)$ term dominates and we solve for
\begin{eqnarray}
     \delta = \frac{1}{2} \exp \Big \lbrace \frac{32 \pi^2}{N_f}M_P^2 \frac{H' M_*}{H^3} \rbrace
\end{eqnarray}
For $\delta$ to be consistently small we need
\begin{equation}\label{PhiCondition}
-\frac{32 \pi^2}{N_f}M_P^2 \frac{H' M_*}{H^3} \gtrsim 1
\end{equation}
When this condition is satisfied, it is consistent to take $\dot{\phi} \simeq H M_*$.

Similarly to the above discussion of the dominant contribution to the relation (\ref{SingularEoM}), we wish to simplify the Friedmann equation $3H^2 = {\cal H}_{eff}/M_P^2$ and work with the most important contribution to the energy density ${\cal H}_{eff}$.  As above, the dominance of the most singular terms as we approach the speed limit is immediate in the regime $\log(1/\delta)\gg 1$.  For applications in which we wish to incorporate early universe phenomenological constraints, we must instead consider the regime $\delta\ll 1$ but $\log(1/\delta)\sim {\cal O}(1)$.  With $\delta\ll 1$, we can again tune the parameters in the non-log terms in the effective Lagrangian in order to ensure that the logarithmic term dominates in the energy density.

In this regime, where (for either of these reasons) the most singular terms dominate the kinetic sector contribution to equation $3H^2 = {\cal H}_{eff}/M_P^2$,  we obtain using (\ref{SingularEoM})
\begin{equation}
3M_P^2 H^2 = V - 2 M_P^2 HH'M_*
\end{equation}
up to terms that are subleading when Equation \ref{PhiCondition} holds. This reduces to the ODE
\begin{equation}\label{Hprime}
H' = \frac{1}{2 M_P^2 H M_*}\big( V - 3M_P^2 H^2 \big)
\end{equation}
which can be integrated to
\begin{equation}\label{Hresult}
H(\phi) = \sqrt{H_0^2 e^{-3(\phi-\phi_0)/M_*} + \frac{1}{M_P^2}\int_{\phi_0}^{\phi} e^{-3(\phi - \phi')/M_*} V(\phi') \frac{d\phi'}{M_*}}
\end{equation}
Note that this doesn't yet give $H(t)$, to do so we'd need to plug this back into 
\begin{equation}
\dot{\phi} = H(\phi)M_*(1-\delta(\phi))
\end{equation}
to obtain $\phi(t)$, and by extension $H(t)$. This equation is not analytically solvable in full generality, though it is numerically tractable. Fortunately, for the sake of checking whether these solutions represent an inflationary universe (and calculating the relevant e-foldings), it is sufficient to work with $H(\phi)$.

Let us first note that inflation requires slow variation of Hubble $H$ in the sense  $-\dot H/ H^2 < 1$.  This, we can see from (\ref{Hprime}) (by multiplying with $\dot \phi \simeq HM_*$),  requires
\begin{equation}\label{InflationLimit}
V > M_P^2 H^2
\end{equation}
At the same time, we want $H' < 0$, which means that $V < 3 M_P^2 H^2$, which gives a decently sized window for $V$ where our solution is controlled. The last thing to check is that the inequality (\ref{PhiCondition}) also holds at the same time, which requires
\begin{equation}\label{PhiIneq}
3 M_P^2 H^2 - V \gg \frac{N_f}{16 \pi^2}H^4
\end{equation}
Combining the above inequalities we obtain a parameter window that yields inflation
\begin{equation}\label{PotentialWindow}
3 M_P^2 H^2 - 100 \frac{N_f}{16\pi^2}H^4> V > M_P^2 H^2
\end{equation}
Here, $100$ is an arbitrary large (but not parametrically large) number that ensures that the ``much greater than" condition of inequality (\ref{PhiIneq}) holds. These inflationary solutions will start at $V(\phi) \approx 3M_P^2 H^2$, and then the potential will decrease until it reaches the lower end of this interval. When this happens, the condition $-\dot H/ H^2 < 1$ is no longer satisfied and inflation ends. Due to the speed limit $\dot{\phi} < HM_*$, it will take a long time for this to happen, and this will yield solutions with a large number of e-foldings.

Let us now estimate the number of e-foldings for a general potential. We will consider a potential $V(\phi)$ and investigate its late time behavior $\phi - \phi_0 \gg M_*$. Most e-foldings happen in this ``late time" regime, so our analysis will be sufficient to parametrically estimate the number of e-foldings. Due to the exponential suppression in the $d\phi'$ integral in (\ref{Hresult}), we can safely extend the interval $(\phi_0, \phi)$ to $(-\infty, \phi)$ (up to corrections that are exponentially suppressed in $(\phi -\phi_0)/M_*$) and re-write the $d\phi'$ integral as 

%({\color{red} The following sum is only a good approximation to the Taylor expantion up to $n\sim 3\frac{\phi-\phi_0}{M_*}$. This can be seen by Taylor expanding $V$ directly in (\ref{Hresult}). Then, we get the sum below by approximating $\Gamma(n)-\Gamma(n, 3\frac{\phi-\phi_0}{M_*}) \approx \Gamma(n)$. We can make a brief comment on this or leave it as is.})

\begin{equation}
\frac{1}{ M_P^2}\int_0^{\infty} e^{-3\phi'/M_*} V(\phi - \phi')\frac{d\phi'}{M_*} = \frac{1}{3 M_P^2}\sum_{n=0}^{\infty} \frac{(-1)^n V^{(n)}(\phi) M_*^n}{3^n}
\end{equation}
as long as $V(\phi)$ has a convergent Taylor expansion. Equation (\ref{Hresult}) can then be rewritten at late times as
\begin{equation}
H^2(\phi) - \frac{V(\phi)}{3 M_P^2} = \frac{1}{3  M_P^2}\sum_{n=1}^{\infty}\frac{(-1)^n V^{(n)}(\phi)M_*^n}{3^n}
\end{equation}
One thing to note right off the bat is that the right hand side is positive as long as $V(\phi)$ is decreasing as a function of $\phi$. This directly follows from the fact that $\int_{0}^{\infty} e^{-3x} f(x) > 0$ for any strictly negative function $f$. In general, the faster that $V(\phi)$ decreases, the bigger $H^2 - V/(3 M_P^2)$ is, and it becomes easy to satisfy \ref{PhiIneq}.

If the potential is slowly varying over scales $\Delta \phi \sim M_*$, then we only need to keep the first derivative term in the above expansion. While this places a constraint on the potentials we can consider, as we will see explicitly this is parametrically much less limiting than the usual slow-roll conditions. This constraint will be necessary in order to get a large number of e-foldings: as we will see shortly, each e-folding reduces $\phi$ by $\sim M_*$, and for (\ref{PotentialWindow}) to hold we cannot let our potential change drastically over a small number of e-foldings. 

As a test case, we take $V(\phi) = A_p |\phi|^p$ with $\phi < 0$ in our regime of interest (our previous analysis was done with the assumption that $\dot \phi > 0$, but an entirely analogous analysis can be done for $\dot \phi < 0$ as well). We can write $V'(\phi) = -pV(\phi)/|\phi|$ and then
\begin{equation}
H^2(\phi) = \frac{V(\phi)}{3M_P^2}\Big(1+ \frac{p M_*}{3 |\phi|}\Big)
\end{equation}
As long as $|\phi| \gg M_*$, the first derivative term is a small correction, and higher order terms are parametrically suppressed. We now use the above expression to examine the window
\begin{equation}
3 M_P^2 H^2 - 100 \frac{N_f}{16\pi^2}H^4 > V > M_P^2 H^2
\end{equation}
The first inequality amounts to
\begin{equation}\label{GeneralPotentialInequality}
V(\phi) \frac{p M_*}{3|\phi|} > 100 N_f \frac{V(\phi)^2}{M_P^4} \rightarrow V(\phi)|\phi| < \frac{p M_* M_P^4}{100 N_f}
\end{equation}
Recall that the factor of 100 is just a sufficiently large number, a factor of 10 would likely be sufficient but 100 gives a conservative estimate. The second inequality is trivially satisfied in the regime where $pM_*/|\phi| \ll 1$. This is only relevant in the last few e-foldings and it doesn't parametrically change the total number.

Using the speed limit $\dot \phi \simeq H M_*$, we compute the number of e-foldings as
\begin{equation}
N_e = \int dt \, H(t) = \int dt \, \frac{\dot{\phi}}{M_*} = \frac{\Delta \phi}{M_* } \approx \frac{|\phi_0|}{M_*}
\end{equation}
where in the final step we assumed $|\phi_0| \gg |\phi_{final}| \gg M_*$ in the regime of interest. To maximize $N_e$, we saturate the inequality (\ref{GeneralPotentialInequality}) by increasing $\phi_0$ as much as possible (and using the relation $V_0 \approx 3M_P^2 H_0^2$), we find
\begin{equation}
N_e \lesssim \frac{p M_P^2}{100 N_f H_0^2}
\end{equation}
We find that it is easy to obtain a very large $N_e$ while remaining within the controlled regime $\dot \phi \simeq HM_*$.

This is a specific domonstration of what we showed in general in \S\ref{sec:inflationgeneral}\ starting with (\ref{ineqM}).  We now turn to phenomenological predictions and observational constraints.  

\section{Phenomenological predictions}

The tilt of the power spectrum and the level of non-Gaussianity in our model are tied to the sound speed of perturbations \cite{Senatore_2016}.  

We can derive this quantity by replacing $\dot\phi^2$ with $(\partial\phi)^2=-g^{\mu\nu}\partial_\mu\phi\partial_\nu \phi=\dot\phi^2-\nabla\phi^2/a(t)^2$ in the action (\ref{Gammashort}) and expanding  $\phi\to\phi+\delta\phi$ 
we can read off the sound speed:  the quadratic terms in $\delta\phi$ take the form $A\delta\dot\phi^2-B(\nabla\phi)^2/a(t)^2$, and $c_s^2=B/A$.  
For our model, near the speed limit this boils down to
\begin{equation}\label{cs}
    c_s^2\simeq -(\frac{HM_*}{\dot\phi}-1)\log(\frac{HM_*}{\dot\phi}-1)\ll 1, ~~~~~~ \dot\phi\approx HM_*
\end{equation}

The current observational constraint on this quantity is \cite{Akrami:2019izv}
\begin{equation}\label{csbound}
    c_s^2\ge 0.021 ~~ (95\% ~ {\text{confidence}})
\end{equation}
which still leaves room for evolution near our speed limit.  Using the formulas detailing our proximity to the speed limit in \S\ref{sec:inflationdetails}, we can trade this for the following constraint on model parameters:  
\begin{equation}
    \gamma\equiv \frac{32\pi^2}{N_f} M_*M_P^2 \frac{|H'|}{H^3} \simeq \frac{32\pi^2}{N_f H^4}(3 M_P^2 H^2-V) \le 2.8
\end{equation}
Incorporating the inequality $N_f\ll \frac{M_P^2}{H^2}$ we find a window of parameters
\begin{equation}\label{VwindowNG}
    3(32\pi^2)(1-\frac{V}{3 M_P^2 H^2})\ll \gamma \ll 2.8 \Rightarrow 1-\frac{V}{3 M_P^2 H^2}\ll \frac{1}{32\pi^2}
\end{equation}
This constraint fits with our inflationary dynamics, in which the potential energy dominates in the Friedmann equation:  as such, $V$ is indeed very close to $3 M_P^2 H^2$.  However, as anticipated in the previous section, this puts the model in the regime with $\delta\ll 1$ but $\log(1/\delta)$ of order 1, necessitating some tuning to fit with early universe observations as described above.\footnote{We have not assessed applications to dark energy, something that involves extremely low energy scales.}  Regardless, the theory of accelerated expansion is of interest for many reasons, including conceptual questions in quantum gravity and the symmetry structure of inflation; it would be interesting to connect to recent works such as \cite{Pajer:2018egx}.   

The bound (\ref{csbound}) derives from observations which have constrained certain shapes of non-Gaussianity \cite{Akrami:2019izv}. Our model is effectively single-field, with inside-horizon interactions supporting inflation, similarly to the case of DBI inflation \cite{Silverstein:2003hf, Alishahiha:2004eh}.  In this situation, we expect equilateral non-Gaussianity \cite{Babich:2004gb}.  

The tilt of the power spectrum is also observationally constrained.  In a general continuous shift-symmetric model such as ours, covered by the simplest version of the EFT of inflation, this is given by (see e.g. \cite{Senatore_2016}\ equation (226))
\begin{equation}
    n_s=1+4\frac{\dot H_*}{H_*^2}-\frac{\ddot H_*}{\dot H_* H_*}-\frac{\dot c_{s*}}{c_{s*}H_*} = 0.9626\pm 0.0057
\end{equation}
where the subscript $_*$ denotes the value of the corresponding quantity at the sound horizon where the mode freezes out:  $\frac{k}{a(t_*)}=\frac{H}{c_s}$.
The current observational constraint \cite{Akrami:2018odb}\ fixes one combination of our parameters within the indicated window.  This is straightforward given the hierarchies established in the previous section. We should note, however, that small corrections that do not affect the inflationary background, e.g. coming from interactions with additional massive fields in a UV completion,  could contribute to the tilt at the observed level.  As such, fixing the particular combination of parameters within our effective field theory is not as meaningful as the above constraint on $c_s$.    

As a last phenomenological comment, we note the current bound on the tensor to scalar ratio,
\begin{equation}\label{rbound}
    r=-16 c_s\frac{\dot H}{H^2} < 0.064 ~~~(95\%~{\text{confidence}})
\end{equation}
Finally, we leave for the future other aspects such as the question of eternal inflation for larger fluctuations, as well as a more systematic study of the parameter space of models in this class.

\noindent{\bf Acknowledgements}

\smallskip

\noindent We are grateful to Victor Gorbenko and David Stefanyszyn for discussions.  
This research was supported in part by the Simons Foundation Origins of the Universe Initiative (modern inflationary cosmology collaboration) and by a Simons Investigator award.

\appendix

\section{Inequalities and parameter windows}

In our analysis in the main text, we imposed various hierarchies and (in)equalities among our scales and parameters.  Here we collect them all in one place and demonstrate their mutual consistency.

\begin{eqnarray}\label{sec:inequalities}
     \dot\phi \simeq M_*H \ll M_{UV}^2\\
     K_\phi = (1+\delta K_\phi)+\frac{N_f}{32\pi^2}\frac{2 M_{UV}^2-H^2}{ M_*^2} \ll \frac{M_P^2}{M_*^2}\\
     \Rightarrow \frac{H}{M_*}\ll\frac{M_{UV}^2}{M_*^2}\ll \frac{64\pi^2}{N_f}\frac{M_P^2}{M_*^2}\\\nonumber
\end{eqnarray}
The first line says we work near the speed limit for $\dot\phi$, and we have specified that the UV cutoff on the $\chi$ mode integral is large compared to this value.  The second line expresses the requirement that the factor $K_\phi = \phi_c^2/\phi^2$ is sufficiently small that the steepness allowed by the condition $(\partial_\phi V M_*)^2/V \ll 1$ for our potential $V(\phi)$ remains an improvement over the slow roll case $(\partial_{\phi_c}V M_P/V)^2 \ll 1$, despite the rescaled kinetic term compared to the canonical field $\phi_c$.    
Combining these leads to the window required for $M_{UV}$ indicated in the third line.   

\section{Full expressions for the action and energy}\label{sec:fullexpressions}

In the main text, we streamlined our analysis at times by invoking the inequalities listed in above in \S\ref{sec:inequalities}\ to simplify some expressions. Here we collect the full expressions for the 1PI effective action and the energy density  (Hamiltonian).  
\begin{align}
    \Gamma_{1PI}=\int a(t)^3\Big\{ \frac{1}{2}K'_\phi\dot\phi^2+\frac{N_f}{64\pi^2}\Big[&\Big(H^4-\frac{\dot \phi^4}{M_*^4}\Big) \log \Big(1-\frac{\dot \phi^2}{M_*^2 H^2}\Big) +\frac{\dot\phi^4}{M_*^4}\log(\frac{M_{UV}^2}{H^2}) - \nonumber\\&
    (M_{UV}^4 - \frac{\dot \phi^4}{M_*^4}) \log \Big(1-\frac{\dot \phi^2}{M_*^2M_{UV}^2}\Big) \Big] -V(\phi)\Big\} 
\end{align}
where
\begin{equation}
K'_\phi = 1+\delta K_\phi+\frac{N_f}{32\pi^2}\frac{ M_{UV}^2-H^2}{ M_*^2}
\end{equation}
\begin{align}
    {\cal H}_{eff}= \frac{1}{2}K'_\phi\dot\phi^2 + \frac{N_f}{64\pi^2}\Big[& -(3\frac{\dot\phi^4}{M_*^4} + H^4)\log \Big(1-\frac{\dot \phi^2}{M_*^2 H^2}\Big)+ 3\frac{\dot\phi^4}{M_*^4}\log(\frac{M_{UV}^2}{H^2}) +
    \nonumber\\&
    2\frac{M_{UV}^2-H^2}{M_*^2}\dot\phi^2+(3\frac{\dot\phi^4}{M_*^4} + M_{UV}^4)\log \Big(1-\frac{\dot \phi^2}{M_*^2M_{UV}^2}\Big)\Big] + V(\phi) 
\end{align}

\bigskip

%%%%%%%%%%%%%%%%%%%%%%%%%%%%%%%%%%%%%%%%%%%%%%%
%%%%%%%%%%%%%%%%%%%%%%%%%%%%%%%%%%%%%%%%%%%%%%%
%%%%%%%%%%%%%%%%%%%%%%%%%%%%%%%%%%%%%%%%%%%%%%%

\bibliography{LogI}{}
\bibliographystyle{utphys}

\end{document}